\documentclass[aps,showpacs,epsf,preprint]{revtex4}
\usepackage{amssymb}
\usepackage{amsmath}
\usepackage{epsf}
\usepackage{mathrsfs}
\usepackage[dvips]{graphics}
\usepackage[dvips,usenames]{color}
\begin{document}
\def \beq{\begin{equation}}
\def \eeq{\end{equation}}
\def \bea{\begin{eqnarray}}
\def \eea{\end{eqnarray}}
\def \bem{\begin{displaymath}}
\def \eem{\end{displaymath}}
\def \f{\frac{\Omega}{\omega}}
\def \fw{\Phi_{\omega}}
\def \fl{\Phi_{\Omega}}
\def \bs{\boldsymbol}
\title{Vortex nucleation through 
edge states in finite Bose-Einstein condensates}
\author{Eric Akkermans and Sankalpa Ghosh}
\affiliation{Physics Department, Technion I.I.T. Haifa-32000, Israel}
\date{\today}

\begin{abstract}
We  study the vortex nucleation in a finite
Bose-Einstein condensate. Using a set of non-local and chiral boundary
conditions to solve the Schr$\ddot{o}$dinger equation of 
non-interacting bosons in a
rotating trap, we obtain a quantitative expression for 
the characteristic angular velocity for vortex nucleation in a condensate 
which is found to be 
$35\%$ of the transverse harmonic trapping frequency. 
\end{abstract}

\maketitle

\section{Introduction}
A prominent feature of a superfluid is the way it behaves
under rotation \cite{Leggett}. 
In contrast to a normal fluid, that rotates like a
rigid body at thermal equilibrium, the thermodynamically stable state
of a superfluid at low enough frequency 
does not rotate. At higher frequencies, 
a finite amount of angular momentum appears in the form of 
vortex filaments at which
the superfluid density vanishes. The circulation of the  velocity field 
flow evaluated on a closed contour that encircles the vortex is
quantized \cite{Feyn,pines}.
This is a consequence of the
existence of a macroscopic wavefunction, whose phase changes
by an integer multiple of $2\pi$ around the vortex filaments.
Atomic Bose-Einstein
condensates (BEC) \cite{book, book1}
provide a reference system where  the superfluid behaviour 
can be studied in the weak-coupling regime.

A set of recent experiments 
has demonstrated  \cite{jila1, jila2, dali1, dali2, 
dali3, mit, mit1, dali4, OXFORD} 
the existence of vortices in atomic condensates.
The 
rotational frequency at which the first vortex is nucleated has been 
shown to depend on parameters such as the trap geometry (aspect ratio),
the nature and the characteristic time of the stirring beam, the number 
of atoms in the trap and may therefore vary from one experiment to another.
The ratio of this characteristic rotational frequency to the transverse  
harmonic trapping frequency varies  
in these experiments from $0.1$ \cite{mit1} to $0.7$\cite{dali2}.

Most of 
theoretical studies of vortex nucleation in atomic BEC
which preceded these experiments
\cite{strin1, lundh, sinha1, 
castin, feder, butt}, determine the characteristic nucleation 
frequency of the first 
vortex from the criterion that 
the vortex-state becomes the minimum of the thermodynamic
free energy of the system evaluated in the co-rotating frame. 
In \cite{IM99} the global as well as local stability
of the vortex state is discussed within the framework of 
the Bogoliubov theory \cite{pines}. 
Using the Thomas-Fermi approximation, the thermodynamic 
characteristic frequency can be expressed \cite{lundh, sinha1}
in terms of the other
system parameters. These works have been reviewed in  
\cite{Fett}. The characteristic rotational frequency 
of the vortex nucleation thus
obtained is generally lower than the values observed experimentally
\cite{dali1, mit1, jila2}.

The presence of a vortex
in a static condensate  
is associated with an extra energy relative to the vortex
free state. 
Using the Thomas-Fermi approximation it can be shown  
that this energy is the highest 
when the vortex is located at the center of the 
trap and decreases monotonically as a function of the distance 
between the center of the vortex
and the center of the trap \cite{Fett, KPSZ01, GG01}. 
Within Thomas-Fermi approximation
(without any boundary correction) it can be shown that 
this energy vanishes at the boundary 
of the system. Thus we can define the energy of a vortex-state $E_{v}$
as a function
of the distance $d$, namely the separation between the vortex and the center
of the trap and $E_v(d)$ has a maximum at $d=0$. 
For an increasing rotational frequency this maximum is shifted
from the center to the boundary of the system \cite{Fett, KPSZ01}
and at rotational frequencies higher than the thermodynamic characteristic
frequency this leads to a surface energy barrier to the nucleation 
of a vortex. This explains why a vortex cannot be nucleated in a trapped 
condensate even though the trap is rotated at 
the thermodynamic characteristic frequency.
The characteristic frequency for the vortex nucleation can be determined
if one knows under what condition and at what rotational frequency 
a state carrying finite angular momentum will be transferred from the surface
to the bulk of the condensate by overcoming the surface energy barrier. 
Theoretical studies in this direction  
have been done in \cite{KPSZ01, 
sstrin, zambelli, ua, strin2, CS01, anglin, Fedichev,
simula}. They are based on an analysis of the collective excitations 
localized at the surface of the condensate, namely the
surface modes \cite{Ono}. These modes   
appear as
shape deformations that carry a finite angular momentum
about the axis of rotation \cite{strin2} and have no radial node.
In a rotated condensate these surface modes are excited and this leads 
to the vortex nucleation.
A generalization of the Landau criterion \cite{book,strin2,Landau} allows to
determine at what rotational frequency, the surface mode corresponding to 
a given angular momentum quantum number is excited and leads to 
the nucleation of a vortex. The characteristic 
nucleation determined 
in this way agrees with the value experimentally observed \cite{anglin}.
Vortices are also nucleated 
by exciting a particular surface mode through a controlled trap deformation
\cite{dali2, OXFORD} and this problem has been theoretically studied
in \cite{KPSZ01, CS01}.

Thus far 
theoretical studies of the problem of vortex nucleation in a 
trapped condensate 
has been conducted  in the framework of interacting bosons 
using various approximation schemes. 
It is also known that the stable state of   
a set of non-interacting bosons in an axisymmetric  
harmonic trap
does not have finite angular momentum along the $z$-direction  
for a rotational frequency less than the trap frequency. 
A finite amount of interaction makes vortex states energetically 
feasible at a lower rotational frequency. In this article we study the 
nucleation of a vortex from the surface to the bulk of a system
by solving the  one-particle Schr$\ddot{o}$dinger equation
with a set of non-local and chiral boundary conditions. To that purpose
we start by discussing 
in section II the role of boundary conditions in a many-body 
problem. In section III, we review the problem of a trapped boson rotating at  
a given frequency
in an infinite plane and  briefly discuss the 
corresponding energy spectrum in a finite disc 
while applying Dirichlet boundary
conditions. In section (IV)
we introduce 
the chiral boundary conditions for rotating bosons in a disc
and we show how the Hilbert space splits into bulk and edge
states. We 
subsequently  analyze the nucleation mechanism, {\it i.e.} 
the transfer of angular momentum form edge to bulk.
The characteristic angular rotation is given in terms of 
the trap frequency at
which the first and then successive vortices  are nucleated in the 
bulk. The variation of the size of the bulk region with increasing 
rotation is discussed and its physical implication is pointed out.

\section{Role of the boundary conditions in a many body problem}
We start with the following hamiltonian 
corresponding to  $N$ interacting bosons of mass $m$ 
\beq H_{mb}=\sum_{i=1}^{N} -\frac{\hbar^2}{2m}\nabla_{i}^2 
+\sum_{i,j} V(|\bs{r}_i-\bs{r}_j|) - E_{0} 
\label{MBH}\eeq
where $E_{0}$ is the ground state energy.
Except for some specific cases, one does not know how to diagonalize this
hamiltonian. Therefore some approximation
schemes must be defined 
whose purpose is to obtain an effective quadratic
hamiltonian. We may consider, for instance, the Feynman description 
\cite{Feyn54} that accounts for the excited states under the form 
\beq \phi(\bs{r}_{1}, \cdots, \bs{r}_{N})=F
\phi_0(\bs{r}_1, \cdots, \bs{r}_N) \eeq where 
$\phi_0$ is the exact but unknown ground state wavefunction such that 
$H_{mb}\phi_0 = 0$, and we assume that 
$F = \sum_{i}^{N}f(\bs{r}_{i})$. The writing of 
$F$ as a sum over one body terms is exact for the 
non-interacting problem. For the interacting case, such 
a decomposition assumes that 
the interaction is adiabatically switched on.
The wavefunction $\phi(\bs{r})$ is obtained by minimizing the energy
\beq E = \frac{\int \phi^{\ast}H_{mb}\phi d^{N} \bs{r}}
{\int |\phi|^2 d^{N} \bs{r}} \eeq
where $d^{N}\bs{r}=d\bs{r}_1\cdots d\bs{r}_N$. 
It can be shown that \cite{Feyn54, Eric1}
the effective energy $E$ can 
be written as
\beq E = -\rho_{0}\frac{\hbar^2}{2m}\int d\bs{r}f^{\ast}(\bs{r})
\nabla^2f(\bs{r})
\eeq
where $\rho_0$ is the ground state density. 
To obtain the corresponding spectrum, we have to impose 
boundary conditions on $f(\bs{r})$. Assuming translational 
invariance, Feynman obtained that  
$E=\frac{\hbar^2}{2m}[\frac{k^2}{S(k)}]$, where $S(k)$ is the structure 
factor. In a trapped condensate the translational invariance 
is broken by the presence of a confinement potential. The function  
$f(\bs{r})$ is related to the order parameter $\Psi(\bs{r})$. 
The choices for  
boundary conditions on the order parameter $\Psi(\bs{r})$ 
is broad as it depends on the nature 
of the confining potential and on the effective one body term generated 
by the interaction. It might be thus possible to take into 
account at least partly the effect of interactions
by solving a linear Schr$\ddot{o}$dinger 
like equation under suitable choices of boundary conditions. However in 
the absence of a specific  relation between such boundary conditions 
and the effective interactions, these choices are generally guided by the 
nature of the problem. 
     
In this  paper we study the vortex nucleation in a confined 
geometry by solving a one-particle Schr$\ddot{o}$dinger equation
for the condensate wavefunction $\Psi(\bs{r})$
with a set of non-local and chiral boundary 
conditions. Such boundary conditions have been proposed in order to 
deal with such non-linear problems \cite{ericlh}. 
These boundary conditions are 
motivated by the following considerations. We know \cite{KPSZ01, 
sstrin, zambelli, ua, strin2, CS01, anglin, Fedichev,
simula}
that the dispersion relation for the surface excitations 
determines the characteristic rotational frequency of vortex 
nucleation. The proposed boundary conditions
are designed in order to provide a clear distinction between the bulk 
and edge states of a two dimensional rotating boson gas. 
The angular momentum quantum numbers of
edge states as we shall see are higher than those  
of bulk states. 
Vortices are then nucleated by transferring a state from the edge to
the bulk Hilbert spaces.

\section{A Rotating two dimensional boson gas}
\subsection{Hamiltonian and  energy spectrum for the infinite plane}
We consider the hamiltonian
of a trapped boson in a two dimensional  domain 
rotating with a uniform
angular frequency $\Omega$ : 

\beq H=\frac{ {\bf p}^2}{2m}+\frac{1}{2}m\omega^2 r^2 -\Omega L_z
\label{hm1} \eeq

We define the vector potential 

\beq  {\bf A}_{f}= \boldsymbol{f} 
\times {\bf r} \eeq
where $\bs{f} = (0,0,f)$. 
Corresponding  pseudo-magnetic fields may be defined by  
\beq {\bf B}_{f} = \boldsymbol{\nabla} \times {\bf A}_{f}
 = 
2\bs{f}\eeq
where  $f$ is either $\omega$ or $\Omega$ so that 
the hamiltonian (\ref{hm1}) rewrites under the two equivalent forms  
\beq H=\frac{1}{2m}({\bf p} - m{\bf A}_{\omega})^2 +(\omega-\Omega) L_z
\label{hm2}\eeq
or 
\beq H= \frac{1}{2m}({\bf p} -m {\bf
A}_{\Omega})^2+\frac{1}{2}m(
\omega^2- \Omega^2)r^2  \label{hm3} \eeq
For $\omega= \Omega$, this hamiltonian 
is the Landau hamiltonian of a charged particle in a 
transverse magnetic field written in the symmetric gauge. 
For this special value 
the centrifugal force just offsets the confinement and hence the
bosonic system becomes unstable.

The eigenfunctions $\Psi_{n,l}$ and 
the eigenvalues $E_{n,l}$ of the hamiltonian for the infinite plane 
are characterized by two integer quantum numbers $n,l$ 
where $n \in {\bf N}, l \in {\bf Z}$.   
We define $b_{\omega}=\frac{m\omega}{\hbar}$ and 
$b_{\Omega}=\frac{m\Omega}{\hbar}$. 
The solutions of the corresponding Schr$\ddot{o}$dinger equation are 
\beq
\Psi_{n,l}(r)=C_{n,l}r^{|l|}e^{il\theta}e^{-\frac{b_{\omega}r^2}{2}}
~_1F_1(a,|l|+1;b_{\omega} r^2) \label{rbwf}
\eeq
where $a=\frac{|l|-\frac{\Omega}{\omega}l+1}{2}-\frac{e}{4}=-n$
with $e=\frac{2E_{n,l}}{\hbar \omega}$. $_1 F_1(a,c;x)$ is the
confluent hypergeometric function \cite{Tricomi,AS}
and $C_{n,l}$ is a normalization constant.
The eigenenergies are 
\beq E_{n,l}=\hbar \omega(2n+|l|-\frac{\Omega}{\omega}l+1) \label{eigv} 
\eeq  
For $l < 0$, the eigenvalues increase with increasing $\Omega$, whereas
for $l > 0$, they decrease with increasing $\Omega$.
For $\frac{\Omega}{\omega} < 1$, the ground state is always characterized
by $n=0, l=0$ so that no vortex is 
nucleated. For $\Omega=\omega$, the energy spectrum is made of  
Landau levels that are degenerate in angular momentum.

The current density in a given eigenstate  is defined by  
\beq  {\bf j} = \frac{\hbar}{2mi}(\Psi_{n,l}^{\ast}
\boldsymbol{\nabla} \Psi_{n,l} - \Psi_{n,l}
\boldsymbol{\nabla} \Psi_{n,l}^{\ast} -
2i\frac{m}{\hbar}{\bf A_{\Omega}} |\Psi_{n,l}|^2)
\label{current} \eeq
Its radial component vanishes while its  
azimuthal component is 
\beq j_{\theta}=\frac{\hbar}{m}
(\frac{l}{r}-\frac{m}{\hbar}\Omega r) |\Psi_{n,l}|^2 \eeq
We define the angular momentum dependent radius $r_{l}$ by 
\beq r_{l} = \sqrt{\frac{l}{b_{\Omega}}} \label{edgerad}\eeq
For a given angular momentum state,  
$j_{\theta}(r)$ is positive for $r < r_{l}$ and hence it 
gives a paramagnetic contribution. It is negative
and diamagnetic
for $r > r_{l}$ 
and it vanishes at $r = r_l$.

\subsection{Spectrum with Dirichlet boundary conditions (DBC)}
We  consider now the problem of a trapped
rotating boson in a disc of radius $R$ imposing 
Dirichlet boundary conditions.  The corresponding 
spectrum is derived in the same way  as for the electron 
in a perpendicular magnetic field \cite{Eric1, Eric11, Eric2, Narevich}.

The Dirichlet boundary condition (DBC)  $\Psi_{n,l}(r=R)=0$,  
is according to (\ref{rbwf}) ~$_1F_1(a,|l|+1;\Phi)=0 $ 
where $\Phi=b_{\omega}R^2$ is equivalent to the magnetic flux. 
The energy spectrum is obtained from the zeroes of the confluent
hypergeometric function $_1F_1$, that can be separated 
into two classes according to the sign of the corresponding 
angular momentum.
For $l \ge 0$,  the energy levels are given by
\beq
_1F_1\big(\frac{l(1-\frac{\Omega}{\omega})+1}{2}-\frac{\varepsilon}{4\Phi},l+1;
\Phi\big )=0 \label{dirichlet1} \eeq
with $\varepsilon=\frac{2mE_{n,l}R^2}{\hbar^2} = e\Phi$. 
For $l <0$ the energy levels are given by
\beq
_1F_1(\frac{-l(1+\frac{\Omega}{\omega})+1}{2}-\frac{\varepsilon}{4\Phi},-l+1;
\Phi)=0 \label{dirichlet2} \eeq
It is important to notice that the equation $_1F_1(a,c;x)=0$ has solutions
only for $a <0$ and, in the interval $-p < a <-p+1$,  it has exactly $p$
real solutions. This eliminates the possibility of 
having for rotating bosons, a ground state given by the lowest
Landau level solution for which
$\frac{\varepsilon}{4\Phi}=\frac{l(1-\frac{\Omega}{\omega})+1}{2}$. 

\begin{figure}[h]
\begin{center}
\vspace*{13pt}
\leavevmode \epsfxsize0.75\columnwidth
\epsfbox{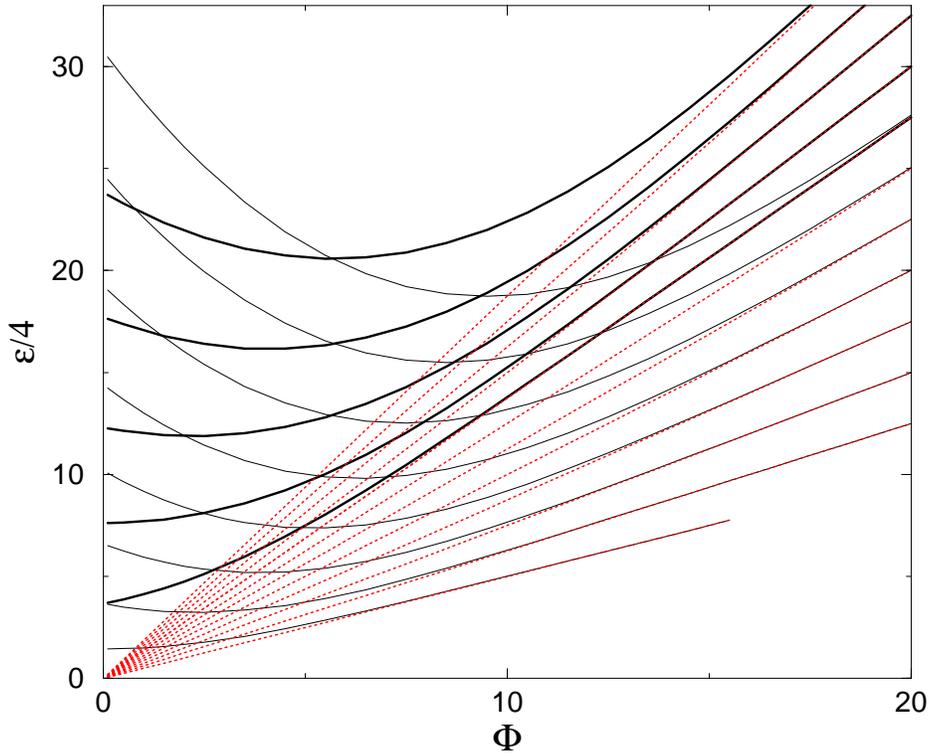}\vskip1.0pc
\caption{\it Energy spectrum of rotating bosons with Dirichlet
boundary condition. The
continuous lines (thin lines for $n=0$ and thick lines for $n=1$)
correspond to the  
energy spectrum derived from (\ref{dirichlet1}).
Each curve corresponds to a given value of the angular momentum.
They  correspond to  $n=0$ and  $l=0, 1, 2, 3, 4, 5, 6, 7$  
as well as $n=1$ and $l=-1, 0, 1, 2, 3$. The dotted lines
represent the corresponding 
infinite plane solutions given by (\ref{eigv}). 
Here $\Omega$ is taken to be $.75\omega$.}
\label{fig1}
\end{center}
\end{figure}

\begin{figure}[h]
\begin{center}
\vspace*{13pt}
\leavevmode \epsfxsize0.75\columnwidth
\epsfbox{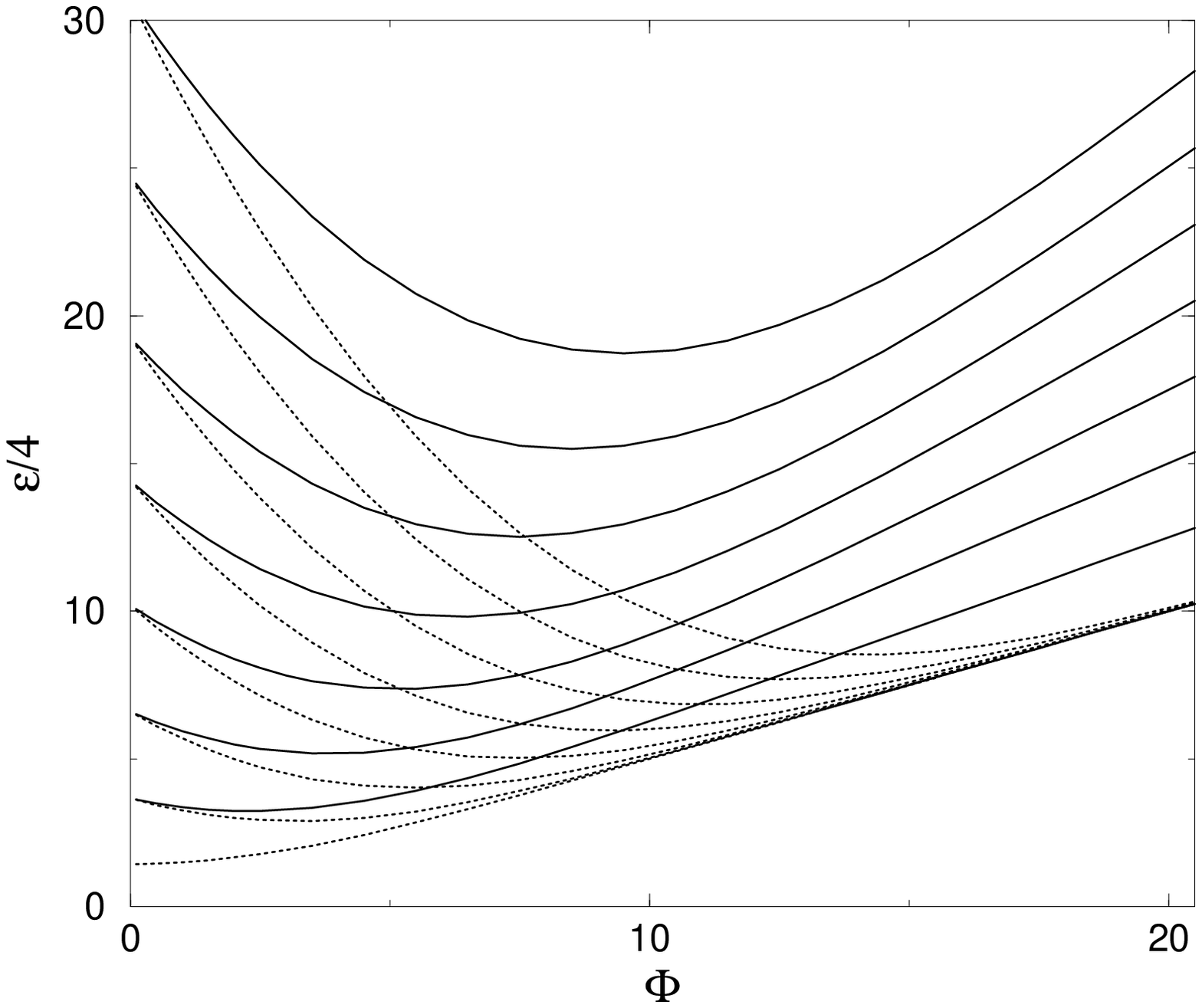}\vskip1.0pc
\caption{\it Comparison between the energy spectrum of 
an electron in a transverse magnetic field (dotted lines) and a
rotating boson under DBC (continuous lines).  
The latter is obtained by setting $\f=1$ in (\ref{dirichlet1}). 
For the curves corresponding to the rotating boson $\f$ is taken to be 
$0.75$. Only the lowest Landau level ($n=0$) is displayed for both cases. 
We have plotted the energy levels for $l$ between $0$ and $7$.
}
\label{fig2}
\end{center}
\end{figure}

The infinite  plane solutions are
reached asymptotically for large $\f$ (see Fig.\ref{fig1}). There 
is no difference in this spectrum between bulk and edge states. 
For $\frac{\Omega}{\omega} < 1$, 
the state $(n,l)=(0,0)$ is always the ground state
(Fig.\ref{fig1}). The energy levels  become degenerate in angular momentum
when $\frac{\Omega}{\omega}=1$ and $\Phi \gg 1$ (Fig. \ref{fig2} and
\cite{Eric1}). 
Therefore  at zero temperature,  no vortex can 
be nucleated at $\frac{\Omega}{\omega} < 1$
under the DBC.

\begin{figure}[h]
\begin{center}
\vspace*{13pt}
\leavevmode \epsfxsize0.75\columnwidth
\epsfbox{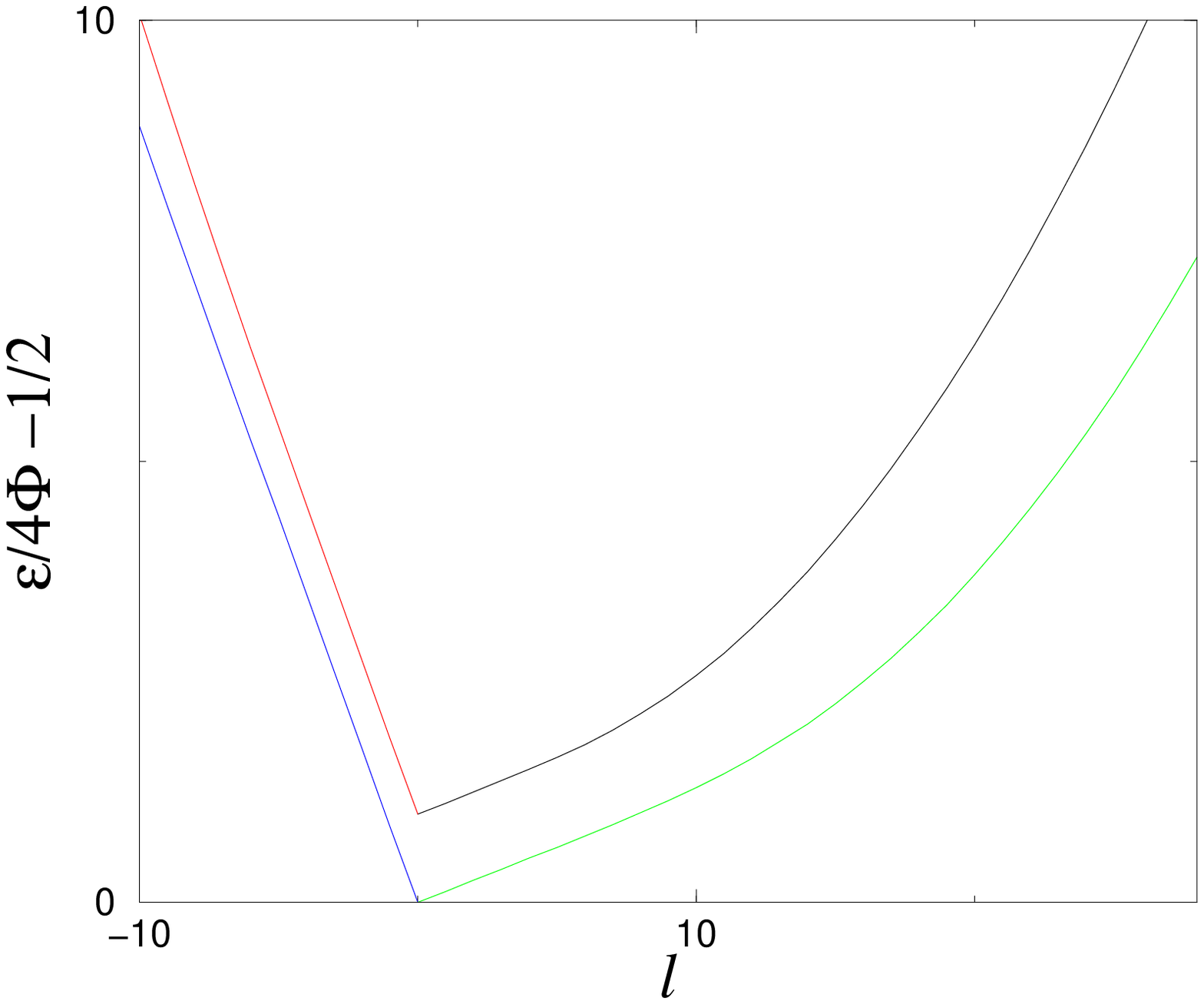}\vskip1.0pc
\caption{\it Edge and bulk states with DBC.
The energy levels that 
correspond to the first two Landau levels are shown.
The lower level corresponds to $n=0$ and the upper one
corresponds to $n=1$. $\Phi$ and the ratio 
$\f$ are respectively taken to $20$ and $0.75$. }
\label{fig3}
\end{center}
\end{figure}
We have  plotted the energy as a 
function of the angular momentum for a fixed $\Phi$ in Fig. \ref{fig3}.
For the Landau problem 
of a charged particle in a transverse magnetic field 
a similar plot, but in a different geometry and only for
positive angular momentum states, has been used in  
\cite{Halperin} in order to describe the
role of the edge states in the quantum Hall transport.
In the present problem as well as in the Landau problem
there is no sharp difference between
edge and bulk states under DBC.

\section{Chiral Boundary conditions}
Dirichlet boundary conditions do not provide
a way to separate edge from bulk excitations.
As already pointed out, such a separation is necessary 
for the description of the nucleation of a vortex. 
We therefore propose a set of non-local and chiral boundary conditions 
which are more suitable for the present problem. 
These boundary conditions are akin to the one 
introduced by Atiyah, Patodi and Singer (APS) in their study  
of Index theorems for Dirac operators with boundaries \cite{APS}.
Similar boundary conditions have also been applied 
to the Landau problem on manifolds with boundaries \cite{Eric2, Narevich}. 
These boundary conditions 
split the Hilbert space into a direct sum of two orthogonal, 
{\it infinite dimensional} spaces with positive and negative chirality
on the boundary. The chirality is determined by the direction
of the azimuthal velocity projected on the boundary.
A vortex
is nucleated as a result of the transfer of a state 
from the edge to the bulk.

We have already noticed that for a given value of the angular momentum
$l$, the current flows with a different chirality in regions 
separated by a ring of radius $r_l=\sqrt{\frac{l}{\fl}}$.
We define the bulk and the edge regions 
using this
particular value of $r_{l}$ as a reference for a given
angular momentum. 
The current associated to that 
particular angular momentum is respectively 
paramagnetic and diamagnetic in  the bulk and at the edge.
Alternatively, we define the bulk and at the edge states for
a disc of size $R$ so that bulk states have angular 
momentum $l < b_{\Omega} R^2$ whereas  edge states have
$l \ge  b_{\Omega} R^2$.

The azimuthal velocity 
 $\frac{j_{\theta}(r)}{|\psi(\vec r)|^2}$, projected on the boundary
of the disc has eigenvalues given by  
\beq  \lambda(R)=\frac{1}{R}(l
-b_{\Omega}R^2)=\frac{1}{R}(l-\Phi_{\Omega}) \eeq
 where 
$\fl=b_{\Omega} R^2$.
The chiral boundary conditions are now defined in the following way:

For $\lambda \ge 0$, namely for $0 < \Phi_{\Omega} \le l$

\beq \partial_r \psi_l|_{R}=0 \label{neumann} \eeq
(For any $n$ and henceforth we shall drop the subscript $n$ in $\psi$.) 

For $\lambda < 0$, namely for $l < \Phi_{\Omega}$

\beq (\frac{\partial}{\partial r}+\frac{i\partial}{r
\partial \theta}+b_{\Omega}r)\Psi_l|_{r=R}= 0 \label{spectral1} \eeq

For the first set of wavefunctions 
that accounts for the edge states we use 
Neumann boundary conditions. 
We could have used as well Dirichlet boundary conditions.
However unlike Neumann boundary conditions they give an
unphysical discontinuity \cite{Eric2}. 
These wavefunctions are more 
and more localized towards the outer 
side of the system for an increasing rotation
frequency.
For states $l < \Phi_{\Omega}$, 
that corresponds to wavefunctions localized well inside the disc, we impose
the mixed boundary condition (\ref{spectral1}). 
It is this separation in the dispersion relations of the edge states 
under the choice of CBC that mimics the effect of interactions. 
In the many-body theory of vortex nucleation
this separation is achieved by solving the 
Bogoliubov equations under various approximations 
\cite{strin2, anglin, simula} and then by 
identifying the collective excitations
localized at the surface of the condensate. 

\subsection{Spectrum with CBC}
\begin{figure}[h]
\begin{center}
\vspace*{13pt}
\leavevmode \epsfxsize0.75\columnwidth
\epsfbox{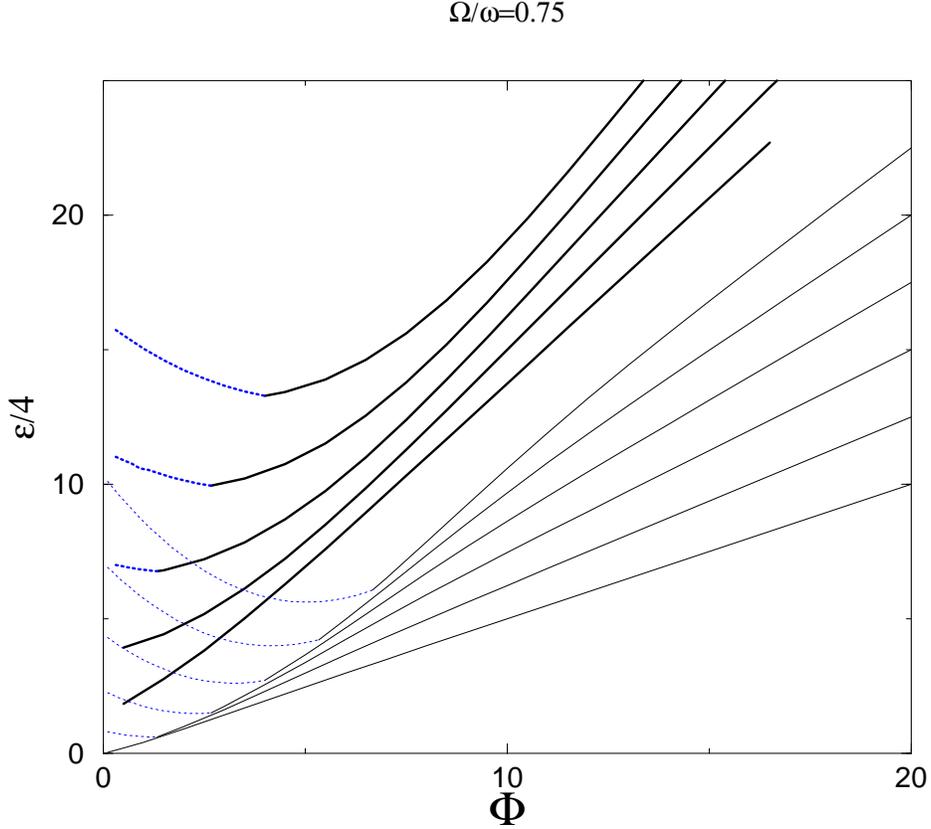}\vskip1.0pc
\caption{\it Energy levels with chiral boundary condition. 
Energy levels for the first few angular momentum
states are shown for $n=0$ and $n=1$. 
Each curve corresponds to a given value of the angular momentum.
They correspond to $n=0$ and  
$l$ between $0$ and $5$ (thin lines for bulk states and 
thin dotted lines for edge states) as well as $n=1$ 
and $l$ between $-1$ and $3$ with $l=-1$
corresponds to the lowest curve (thick lines for bulk states
and thick dotted lines for the edge states). 
For $l > 0$ 
the spectrum has
a kink at the point $l=\Phi_{\Omega} =\f \Phi$.}
\label{fig4}
\end{center}
\end{figure}

According to the chiral boundary conditions when $\Phi$ is increased
at a fixed $\f$, the sign of the eigenvalues 
$\lambda(R)$ changes from  the positive to 
the negative. Correspondingly the energy $\varepsilon$  of 
a state with a given $n$ and $l$ changes. This change in energy 
describes how the corresponding state is  
transferred from the edge Hilbert space to the bulk Hilbert space at the point 
$\Phi_{\Omega}=l$ (\ref{neumann}).
This is shown in Fig.\ref{fig4}.   
For large $\Phi$, the infinite plane solutions (\ref{eigv}) are reached
asymptotically.  

\begin{figure}[h]
\begin{center}
\vspace*{13pt}
\leavevmode \epsfxsize0.75\columnwidth
\epsfbox{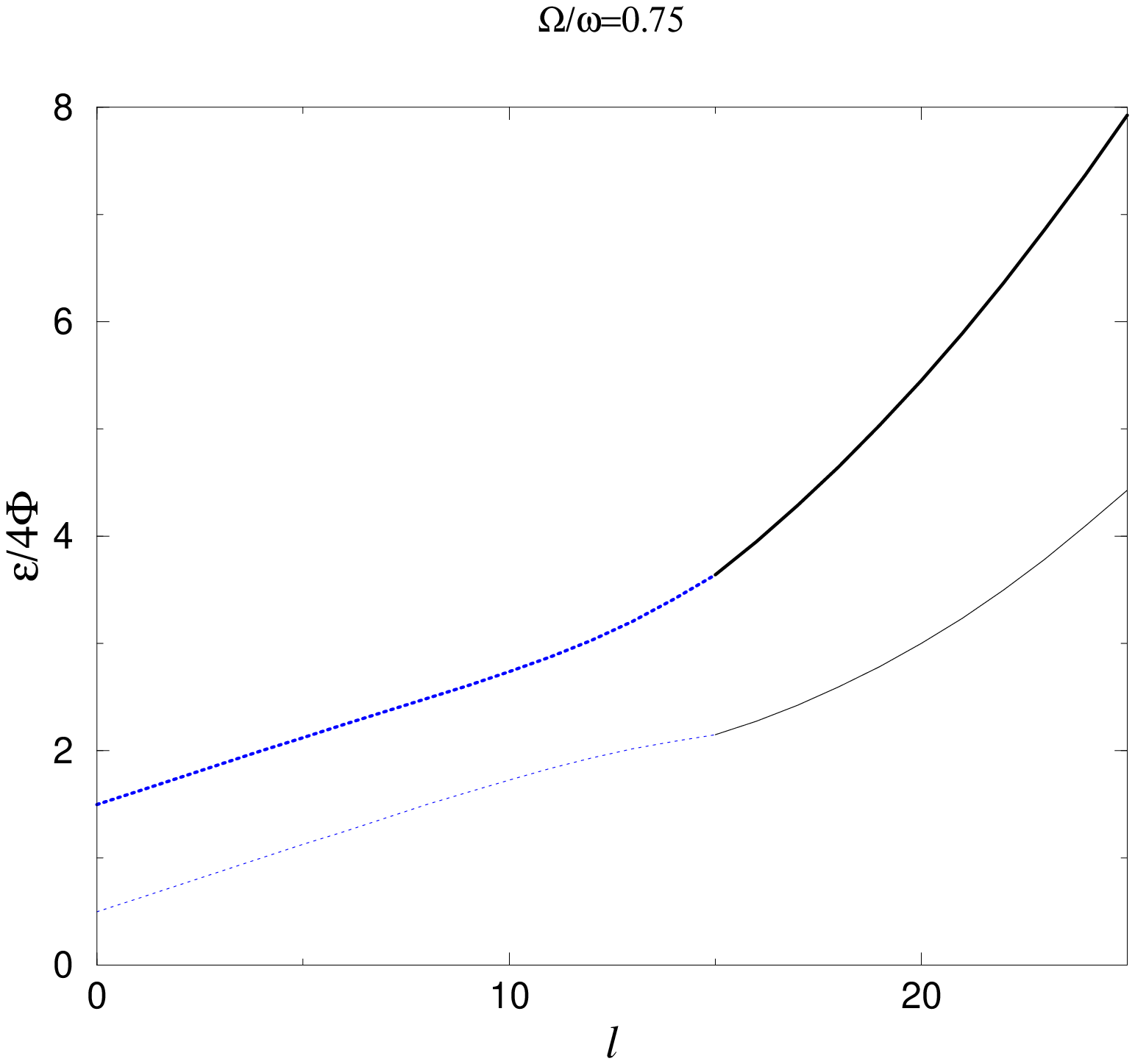}\vskip1.0pc
\caption{\it Edge and bulk states with chiral 
boundary conditions. 
Energy levels that correspond to  $n=0$(lower curve) and $n=1$ 
(upper curve) are shown. Here the edge states are denoted
by continuous lines whereas the bulk states are 
denoted by dotted lines.
The slope of the energy levels is discontinuous at
$l=\Phi_{\Omega}$ where $\Phi_{\Omega}$ is taken to be 15.}
\label{fig5}
\end{center}
\end{figure}

For an infinite system we have $\frac{\varepsilon}{4\Phi} = 
n+\frac{1}{2}(1+(1-\f)l)$. 
Therefore $\frac{\varepsilon}{4\Phi}$, for a given $\f$, 
is a linear function of $l$ with slope $(1-\f)$.
When chiral boundary conditions are applied, 
this behaviour is approximately obeyed for
the bulk states. But for the edge states the energy increases
non-linearly with increasing angular momentum. This is shown in 
Fig.\ref{fig5}. 
This is also the case for Dirichlet boundary conditions (Fig.\ref{fig3}). 
There is however a quantitative
difference and moreover the bulk and the edge states are now separated
under these chiral boundary conditions.
For a given $n$ 
the spectrum is continuous at the point $l=\Phi_{\Omega}$, 
but its derivative is not. This has been shown in 
Fig.\ref{fig5} for the first two values of $n$. 
The derivation
of the spectrum under
these boundary conditions is provided in detail 
in the appendix. Here we focus on the 
nucleation of vortices under these boundary conditions.

\begin{figure}[h]
\begin{center}
\vspace*{13pt}
\leavevmode \epsfxsize0.75\columnwidth
\epsfbox{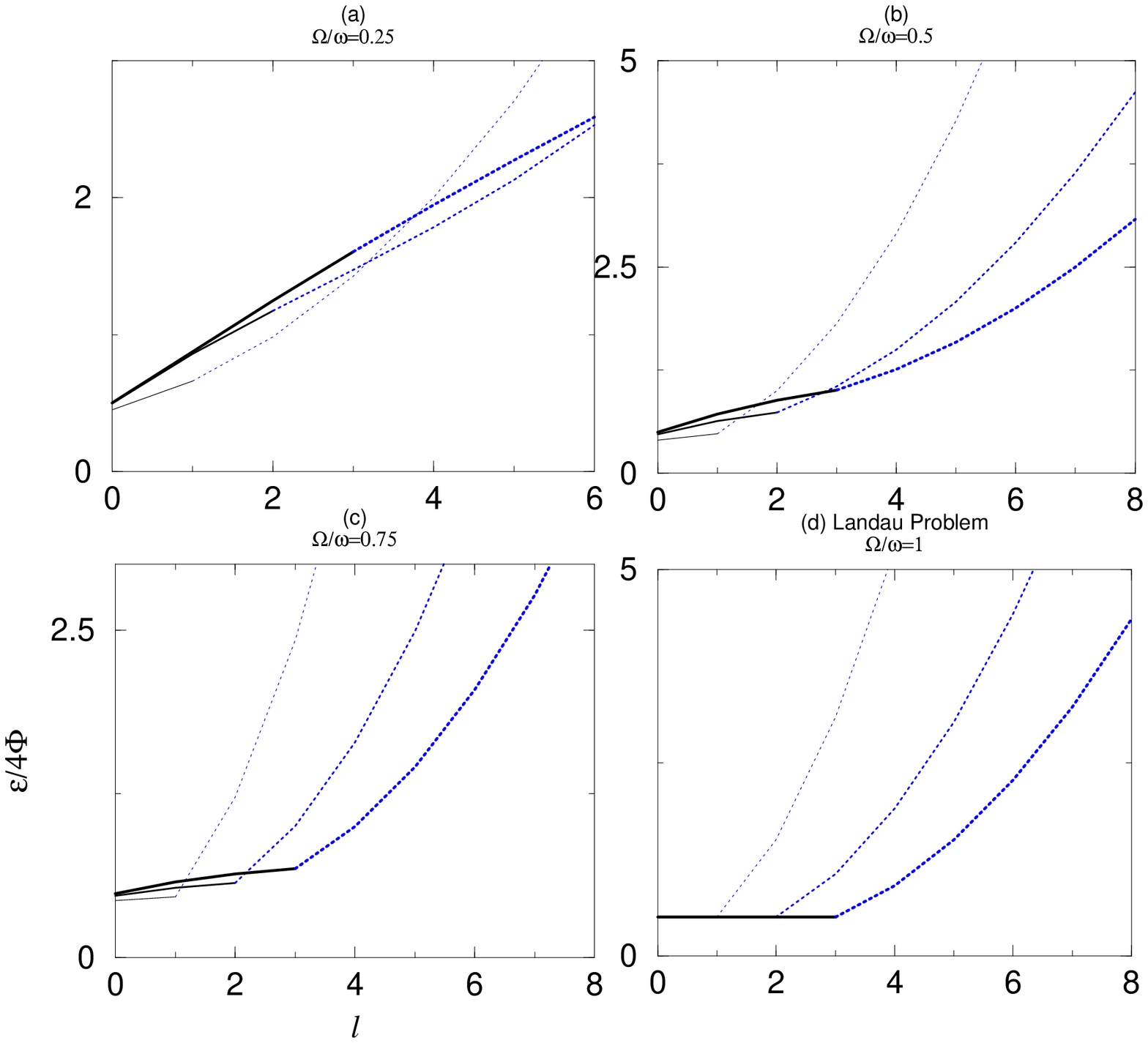}\vskip1.0pc
\caption{\it Effect of a faster rotation.
In these figures we have plotted 
the energies $\frac{\varepsilon}{4\Phi}$ 
as a function of $l$ for a set of $\f$ values (given above each figure). 
The three  plots in each figure correspond to $\Phi_{\Omega}=1,2,3$
(the thinnest one for $\fl=1$ and the thickest for $\fl=3$). 
The dotted part corresponds to edge states while the continuous part
corresponds to bulk states. For $\f=1$ bulk states for all three 
values of $\fl$ fall on the same line.}
\label{fig6}
\end{center}
\end{figure}

\begin{figure}[h]
\begin{center}
\vspace*{13pt}
\leavevmode \epsfxsize0.75\columnwidth
\epsfbox{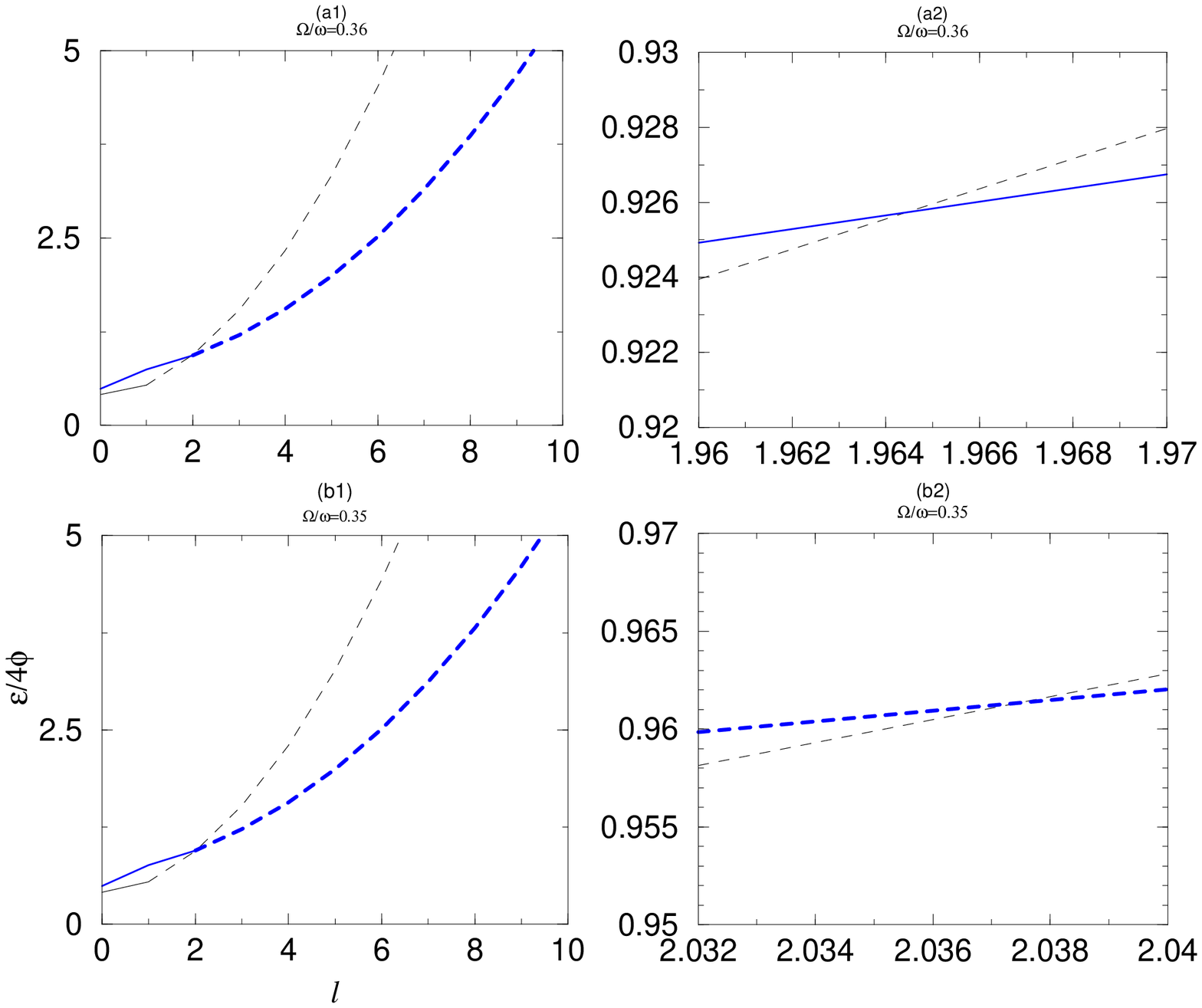}\vskip1.0pc
\caption{\it Determination of the characteristic  frequency
of vortex nucleation.
In these figures we have plotted 
the energy $\frac{\varepsilon}{4\Phi}$ 
against the angular momentum $l$.
The two curves 
in each figure correspond to $\Phi_{\Omega}=1$ (thin), $\fl=2$(thick). 
Corresponding ratios $\frac{\Omega}{\omega}$ are mentioned 
above each figure. In each curve 
the continuous part corresponds to bulk states and the
dashed-line part to edge states. 
At $\f=0.35$ the edge spectrum
for $\fl=1$ intersects the corresponding edge spectrum for $\fl=2$
($b1$ and $b2$). 
At $\f=0.36$, the intersection takes place between the bulk spectrum
for $\fl=2$ and the edge spectrum for $\fl=1$ ($a1$ and $a2$).  
The regions near points of intersection  are enlarged  in the
figures $a2$ and $b2$.}
\label{fig7}
\end{center}
\end{figure}

The Fig.\ref{fig6} shows the effect
of an increase of the  rotational frequency 
on the spectrum under the choice of CBC.  
We have plotted 
the energies of the bulk and the edge states 
for four different values
of the ratio $\f$. For each value, 
the quantity $\Phi_{\Omega}$ 
is increased by unit steps from
$1$ to $3$ and the corresponding bulk and edge energies are shown.
Under these conditions, 
the slope of the bulk energy levels increases while the slope
of the edge energy levels goes down.
The opposite behaviour is observed when, for a fixed $\fl$, the ratio $\f$
increases. 
Therefore with increasing rotational frequency, states with
higher angular momenta are transferred from the edge Hilbert space to
the bulk Hilbert space. The $l$-th angular momentum state is nucleated
in the bulk from the edge by changing 
$\fl$ from $l$ to $l+1$. 

The plots that appear in Fig.\ref{fig6} represent solutions 
of the stationary Schr$\ddot{o}$dinger equation under the choice of CBC
at different values of $\Omega$ and $\fl$. 
To understand the nucleation of vortices in the condensate with these
solutions we use the fact that a condensate is rotated only by 
nucleating a vortex. Therefore in between nucleations of 
successive vortices, the radius of the condensate remains constant. This
is in contrast to the rotation of a rigid body which flattens out 
continuously while increasing the rotational frequency. Let us denote by 
$\Omega_1$, the characteristic frequency for nucleation of the first vortex.
For a given 
$\Omega < \Omega_1$, the radius of the condensate is 
$R$ and it corresponds to $\fl =1$. The 
corresponding bulk and edge energy levels for $\fl=1$ 
are obtained from (\ref{neumann}) and (\ref{spectral1}). The condensate
is therefore the bulk region that contains only the $l=0$ state.
At $\Omega
= \Omega_{1}$, the first vortex is nucleated by transferring the $l=1$ state
from the edge to the bulk so that 
$\fl$ changes from $1$ to $2$. 
Therefore the condensate is now defined as 
the bulk region of $\fl=2$ 
and it has a larger radius. The rotational flux transferred by 
nucleating a vortex
is thus $\frac{h}{m}$ since this is the  unit of $\fl$.
At $\Omega= \Omega_1$, the edge states for $\fl=1$  intersects 
the bulk states for $\fl=2$. This is energetically favoured since 
at $\Omega =\Omega_{1}$, the energy 
$\frac{\varepsilon}{4\Phi}(l)$  of any state with $l \ge 2 $
is less if  $\fl=2$ rather than $\fl=1$. 
In the corresponding many body description 
\cite{KPSZ01, anglin, simula}, vortices are nucleated 
when at higher rotational frequency, a surface energy barrier
disappears. The reduction of the edge state energy with the nucleation of 
a vortex is qualitatively similar to the disappearance of the surface 
energy barrier.   

The characteristic rotational frequency for the nucleation of the first
vortex determined using chiral 
boundary conditions (\ref{neumann},\ref{spectral1}) is   
between $\Omega=0.35~\omega$ and $0.36~\omega$ as shown in Fig.\ref{fig7}
where $\omega$ is the transverse trap frequency.  
We note
that this characteristic nucleation frequency is close to the value
$\Omega=
0.29~\omega$  that has been observed in one 
of the experiments on vortex nucleation \cite{mit}. 
All these features cannot be derived from either the infinite
plane or from DBC.

To establish a parallel with the case of superconductors 
let us mention similar results obtained for the vortex nucleation
in a mesoscopic superconducting disc \cite{Peeters, Eric3, Lopez}. 
However in a neutral superfluid there is no Maxwell-Amp$\grave{e}$re 
equation that can relate the current to the induced rotation .

\section{Vortex nucleation for $\Omega > \Omega_1$}
Thus far we have discussed the nucleation of the first vortex.
Here we discuss the bulk and edge spectrum under CBC for $\Omega > \Omega_1$.
We consider the following two cases:
\begin{enumerate}
\item
For a further increase of  $\Omega$ beyond $\Omega_{1}$, the edge states 
corresponding to 
$\fl=l$ crosses the bulk states  
of $\fl=l+1$ for $l \ge 2$
(Fig.\ref{fig6}). The corresponding rotational frequencies 
$\Omega_{l}$ are the nucleation 
frequencies of the $l$-th vortices. 
Each of these vortices is nucleated 
by changing $\fl$ to $\fl +1$ and it carries one unit of angular momentum.
Characteristic rotational frequencies
are presented in the Table \ref{one-vortex}. 
Since the bulk and edge energy spectra under CBC 
is obtained by solving the stationary Schr$\ddot{o}$dinger equation 
in a co-rotating frame, solutions
at different values of $\Omega$ can be related one to the  other if 
the rotation is switched on adiabatically.
Such experiments \cite{dali2, OXFORD} 
have been performed 
in a deformed rotating trap where the cross-section of the 
rotating condensate is an ellipse in the plane of rotation. 
Here we solve  
the Schr$\ddot{o}$dinger equation 
in a circular domain. This 
makes the comparison with the experimental results rather difficult.
The theoretical work \cite{CS01} used an approach 
different from ours, to explain the nucleation
mechanism of the first vortex. It should also be noticed that   
the adiabatic nucleation of successive vortices has not yet been
observed experimentally. For a recent theoretical work on this issue
see \cite{LSC04}. 
One of the limitations of our theoretical frame-work
is that we are unable to determine the spatial 
arrangement of more than one vortex inside the condensate. 

\begin{table}[tbp]
\begin{tabular}{|c|c|}
\hline
Vortex & $\f$\\
\hline
$1$st &  $0.35$ - $0.36$\\
$2$nd & $0.46$-$0.47$ \\
$3$rd &  $0.52$-$0.53$ \\
$4$th &  $0.569$-$0.57$ \\
$5$th &  $0.603-0.604$ \\
$6$th &  $0.629-0.630$ \\
\hline
\end{tabular}
\caption{\it Nucleation frequencies of successive ($l=1$) 
vortices.  
}
\label{one-vortex}
\end{table}

\item Another interesting point is to find 
the rotational frequencies at which
edge states of $\fl=1$ intersects the bulk state $l=2$ for   
$\fl=3, 4, \cdots$.  The bulk and edge energy levels of the stationary 
Schr$\ddot{o}$dinger equations at these rotational frequencies
correspond to a situation 
where the rotation is suddenly switched on at a frequency 
higher than $\Omega_{1}$ and subsequently the  system is  
brought to equilibrium in the co-rotating frame. 
More than one 
angular momentum states (and rotational fluxes) 
are transferred to the bulk at these characteristic 
frequencies. This leads to the nucleation of a vortex of 
angular momentum $l > 1$ or multiple $l=1$ vortices  
at a time. Within
the present theoretical 
framework it is not possible to identify which of these two 
alternatives is energetically 
favourable.  In Table \ref{many-vortex}
we have listed these rotational frequencies and the number of 
angular momenta transferred to the bulk Hilbert space. 
\begin{table}[tbp]
\begin{tabular}{|c|c|}
\hline
No. of   & $\f$\\
states transferred  & \\
to the bulk & \\
\hline
$2~(1,2)$  &  $0.44$ - $0.45$\\
$3~(1,2,3)$  & $0.48$-$0.49$ \\
$4~(1,2,3,4)$  &  $0.494$-$0.495$ \\
$5~(1,2,3,4, 5)$  &  $0.498$-$0.499$ \\
\hline
\end{tabular}
\caption{\it Nucleation frequencies of multiple vortices. In parentheses
we mention the angular momentum states which are transferred from
the edge to the bulk at these values of $\Omega$.}
\label{many-vortex}
\end{table}
Around $\f = 0.5$, the number of vortices nucleated in this way increases 
very fast.
Experimentally \cite{dali3, mit} it has been observed that 
under sudden switch-on of the rotation at higher rotational  
frequencies more than one vortices are nucleated and  
the number of nucleated vortices is the largest,
(see for example Fig. 3 in \cite{mit})
at $\Omega=\frac{\omega}{\sqrt{2}}$. This is associated with a resonant
quadrupolar excitation of angular momentum $l=2$. 
The energy of this 
quadrupolar surface mode in units of the transverse trap frequency
is determined within the classical hydrodynamic
approximation that is very different from our approach. Therefore 
although our findings
qualitatively agree with these experimental observations, 
a quantitative comparison with both the experiments and the related theory
is rather involved. A reason for that is 
the difference in the trap geometry in the two-dimensional plane
which is elliptical in the experiment while circular in our case. 
Another reason is the different theoretical approach that have been  
used to study the surface excitations in the many-body theory
\cite{ua, strin2, anglin} and our. The difference 
between the characteristic rotational frequencies in the above two cases 
(Table \ref{one-vortex} and \ref{many-vortex}) thus  
shows that the nucleation frequency of a vortex in the 
condensate depends on the number of vortices already 
present. 
\end{enumerate}

\begin{figure}[h]
\begin{center}
\vspace*{13pt}
\leavevmode \epsfxsize0.75\columnwidth
\epsfbox{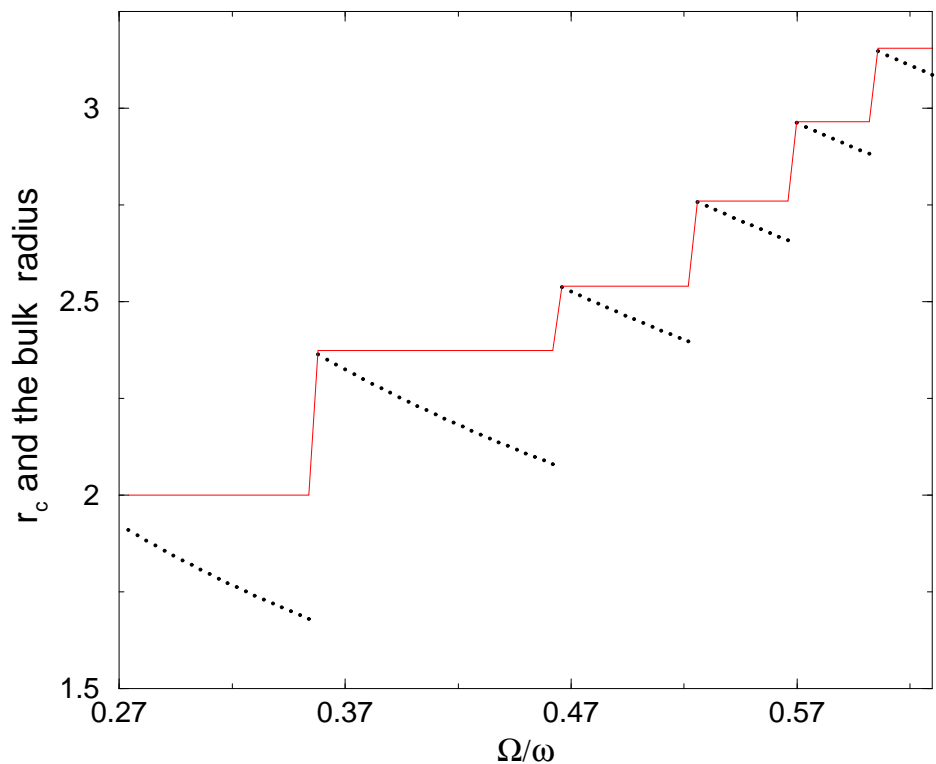}\vskip1.0pc
\caption{\it Size of the condensate as a function of 
the rotational
frequency.
The continuous lines 
gives the radius
of the condensate $r_c$  as a function of $\f$.
The dots give the radius of the bulk
region for $\fl=l$ and for $\Omega_{l} < \Omega < \Omega_{l-1}$. 
The length is measured in units of 
$\frac{1}{\sqrt{b_{\omega}}}$ where $b_{\omega}=\frac{m\omega}{\hbar}$.   
}
\label{fig8}
\end{center}
\end{figure}

The radius of the condensate is changed when a vortex is nucleated. Since 
a boson in the condensate has one more unit of angular momentum
when a vortex is nucleated, its orbit has a 
larger radius. This explains why size of the condensate region increases 
with successive nucleations.
When the $l$-th vortex with 
one unit of angular momentum is nucleated, the condensate region 
corresponds to $\fl=l+1$. Therefore
at each $\Omega_{l}$, the radius of the condensate region 
is defined as 
\beq r_{c}|_{\Omega=\Omega_{l}} = \sqrt{\frac{\hbar(l+1)}{m \Omega_{l}}} \eeq
Between the nucleations of the
$l$-th and $(l+1)$-th vortex, the radius of the bulk region for $\fl=l$
decreases as $r=
\sqrt{\frac{\hbar (l+1)}{m\Omega}}$ and the corresponding edge region 
grows in size. 
In Fig.\ref{fig8} using the set of frequencies 
listed in Table \ref{one-vortex} we have plotted the condensate 
radius. 
In the same figure we have simultaneously 
plotted the radius of the bulk region for $\fl=l$, 
in between the nucleation of the $l$-th and $(l-1)$-th vortices.
This plot involves a set of disconnected lines and the 
corresponding differences
with the condensate region show how the edge region grows 
between the nucleation of two successive vortices.
With the increase of the number of vortices in the condensate these two plots
get closer one to the other. In the  
limit of a large number of vortices, 
the average rotation approaches the rigid body value 
\cite{Feyn, dali3, Fett}. 
For example, it has been shown in \cite{dali3} that 
experimentally this happens 
when the number of vortices is around $10$ \cite{dali3}. 

Khawaja {\it et. al.} \cite{ua} have studied the surface excitations 
of a trapped three-dimensional Bose-Einstein condensed gas 
using the Gross-Pitaevskii equation 
and they have associated the kinetic energy 
of  the surface region to  an effective
surface tension. Although our approach is 
different and that we consider a strictly two dimensional system, 
we can also define an energy 
associated to the edge by summing the energies of the 
edge states for $\fl=l$. 
For a finite system then a part of this edge energy is used 
to nucleate vortices .

\section{Conclusion}
We have presented a one-particle effective theory
that describes the process of vortex nucleation in a
two dimensional rotating condensate 
using a set of non local and chiral boundary conditions.
We have shown that the edge states have a 
dispersion relation distinct from 
the bulk states. These boundary 
conditions can be understood as a way to mimic 
the effect of the interaction between the bosons. 
We have demonstrated that 
the properties of vortices thus nucleated
agree qualitatively and quantitatively 
with experimental findings although a  more thorough  
comparison is not possible in the absence of a precise mapping 
between this set of 
boundary conditions and the effective interaction. 
The expression of the characteristic rotational frequency
at which vortices are nucleated under adiabatic and sudden switch-on of 
the rotation is calculated and the  
change  in the condensate size with increasing rotation is also emphasized. 

\vskip 20pt
\noindent{\large\bf Acknowledgements}\\\\

We thank the Israel Council for Higher Education for financial support.
This work is supported in part by a grant from the Israel Academy of 
Sciences and the fund for promotion of research at the Technion.

\newpage
\noindent{\Large\bf Appendix I}\label{app1}\\
\vskip 10pt
\begin{itemize}
\item {\it Case 1}: 
For
$\lambda \ge 0  \Rightarrow  0 < \Phi_{\Omega} \le l$
plugging the  explicit form of the wavefunction (\ref{rbwf})
in equation (\ref{neumann}) we get \cite{AS}
\beq (\frac{l}{\Phi}-1)_1F_1(a,l+1;\Phi)+\frac{2a}{l+1}~ _1F_1(a+1,l+2
;\Phi)=0 \label{apeq1} \eeq
\item {\it Case 2}: 
For $\lambda < 0$
there are two possibilities \\
\item {\it Case 2a}:  For $l < 0$ ( negative angular momentum)
the spectral boundary condition (\ref{spectral1}) implies
\beq (\frac{\partial}{\partial r}+\frac{i\partial}{r
\partial \theta}+b_{\Omega}r) \Psi_{n,l}|_{r=R} = 0 \label{mixed} \eeq
Using the explicit form of wavefunction given in
(\ref{rbwf}) we find that this demands
\beq 2l~ _1F_1(
a,|l|;\Phi) - \Phi(1-\frac{\Omega}{\omega}) _1F_1(
a,|l|+1;\Phi)=0 \label{apeq2}\eeq
\item {\it Case 2b}: 
The other situation is where $0 \le l <\Phi_{\Omega}$.
Again using the explicit form of the wavefunction ($l > 0$) 
it can be found that
\beq \frac{2a}{l+1}~_1F_1(a+1,l+2;\Phi)-(1-\frac{\Omega}{\omega})_1F_1
(a,l+1;\Phi)=0 \label{apeq3}\eeq
\item
It can be seen that the lowest Landau level solution exists
only if $\Omega=\omega$. If we set $2b_{\omega}=b,\omega=\Omega$
where $b=\frac{m \omega_c}{\hbar}$ and $\omega_c$ is the cyclotron
frequency, all the above results matches with the
results obtained for the Landau problem on a finite disk
under chiral boundary condition \cite{Eric2}.
We solve all these equation with {\it Mathematica} \cite{Math}

\end{itemize}

To obtain the characteristic frequencies for the nucleation of successive
vortices (Table \ref{one-vortex}) we have solved the equation (\ref{apeq1})
for $\fl=l$ and $\fl=l+1$ simultaneously. For $l=1$
the solution yields $0.35 \omega < \Omega < .36\omega$ (Fig. \ref{fig7}). 
The characteristic frequencies listed in Table \ref{many-vortex} are 
obtained from simultaneously solving equation (\ref{apeq1}) with $\fl =1$
and equation (\ref{apeq3}) with $\fl=3, 4, 5, \cdots$.


\begin{thebibliography}{99}
\bibitem{Leggett} A.J. Legget, Physica Fennica {\bf 8}, 125, (1973).


\bibitem{Feyn}R.P. Feynman, Application of Quantum Mechanics to
Liquid Helium in {\it Progress in Low Temperature Physics I},  
Edited by C. J. Gorter (North-Holland Publishing Co., Amsterdam, 1955), 
Chapter 2.

\bibitem{pines}P. Nozieres and D. Pines, 
{\it The Theory of Quantum Liquids, Vol II}, 
(Addition -Wesley Publishing Company, Inc., 1990).

\bibitem{book} C. Pethick and H. Smith, Cambridge University Press,
{\it Bose-Einstein Condensation in Dilute
gases}, (Cambridge University Press, 2002).
For a discussion of the {\it Landau Criterion} see the Chapter 10.
\bibitem{book1}L. Pitaevskii and S. Stringari, 
{\it Bose-Einstein Condensation}, 
(Oxford Science Publication, 2003). 
 
\bibitem{jila1}M.R.Mathews, B.P. Anderson, P.C. Haljan, D.S. Hall,
C. E. Weiman  and
E. A. Cornell, Phys. Rev. Lett. {\bf 83 }, 2498, (1999).

\bibitem{jila2}P.C. Haljan, I. Coddington, P. Engels and E.A. Cornell
, Phys. Rev. Lett. {\bf 87}, 210403, (2001).

\bibitem{dali1}K.W.Madison, F. Chevy, V. Bretin and J. Dalibard,
Phys. Rev. Lett. {\bf 84}, 806, (2000); F. Chevy, K.W.Madison and
J. Dalibard, Phys. Rev. Lett. {\bf 85}, 2223, (2000).

\bibitem{dali2} K.W.Madison, F. Chevy, V. Bretin and J. Dalibard,
Phys. Rev. Lett. {\bf 86}, 4443, (2001).

\bibitem{dali3}F. Chevy, K.W. Madison, V. Bretin and J. Dalibard, 
arXiv:cond-mat/0104218. 

\bibitem{mit} J. R. Abo Shaeer, C. Raman, J.M. Vogels and W. Ketterle,
Science {\bf 292}, 476, (2001).

\bibitem{mit1}C. R. Raman, J. R. Abo-Shaeer, 
J. M. Vogels, K. Xu and W. Ketterle, Phys. 
Rev. Lett. {\bf 87}, 210402, (2001).

\bibitem{dali4}P. Rosenbusch, V. Bretin and J. Dalibard, Phys. Rev. Lett. 
{\bf 89}, 200403, (2002).


\bibitem{OXFORD}E. Hodby, G. Hechenblaikner, S. A. Hopkins, 
O. M. Marag\`{o}, and C. J. Foot, 
Phys. Rev. Lett. {\bf 88}, 010405 (2002).

\bibitem{strin1}F. Dalfovo and S. Stringari, Phys. Rev. A {\bf 53}, 
2477, (1996).

\bibitem{lundh} E. Lundh, C. J. Pethick and H. Smith, Phys. Rev. A {\bf
55}, 2126, (1997).
 
\bibitem{sinha1} S. Sinha, Phys. Rev. A {\bf 55}, 4325, (1997).

\bibitem{castin}Y. Castin and R. Dum, Eur. Phys. J. D {\bf 7}, 399 (1999).

\bibitem{feder} D. L. Feder, C. W. Clark and B. I. Schneider, 
Phys. Rev. Lett. {\bf 82}, 4956, (1999).

\bibitem{butt} D. A. Butts and D. S. Rokshar, Nature {\bf 397}, 327,
(1999).

\bibitem{IM99}T. Isoshima and K. Machida, Phys. Rev. A, 
{\bf 60}, 3313, (1999). 

\bibitem{Fett} A. L. Fetter and A. A. Svidzinsky, J. Phys. C 
{\bf 13}, R135, (2001).


\bibitem{KPSZ01}M. Kraemer, L. Pitaevskii, S. Stringari and F. Zambelli, 
Laser Physics {\bf 12}(1), 113 (2002).


\bibitem{GG01}M. Guilleumas and R. Graham, Phys. Rev. A, {\bf 64}, 033607 
(2001)

\bibitem{sstrin}S. Stringari, Phys. Rev. Lett. {\bf 77}, 2360, 
(1996).

\bibitem{zambelli} A. Recati, F. Zambelli and S. Stringari, Phys. Rev. Lett. 
{\bf 86}, 377, (2001).

\bibitem{ua}  U. A. Khawaja, C. J. Pethick, and H. Smith, Phys.
Rev. A {\bf 60}, 1507, (1999).

\bibitem{strin2}F. Dalfovo and S. Stringari, Phys. Rev. A {\bf 63},
011601(R), (2000).

\bibitem{CS01}S. Sinha and Y. Castin, Phys. Rev. Lett. {\bf 87}, 
190402 (2001)

\bibitem{anglin} J. R. Anglin, Phys. Rev. Lett. {\bf 87}, 240401,
(2001), {\it ibid} Phys. Rev. A {\bf 65}, 063611, (2002).

\bibitem{Fedichev}A. E. Muryshev and 
P. O. Fedichev, cond-mat/0106462.

\bibitem{simula} T. P. Simula, S. M. M. Virtanen and M. M. Saloma, 
Phys. Rev. A {\bf 66}, 035601, (2002).


\bibitem{Ono}R. Onofrio, D. S. Durfee, C. Raman, M. K$\ddot{o}$hl, 
C. E. Kuklewicz and W. Ketterle, Phys. Rev. Lett. {\bf 84}, 810, (2000).


\bibitem{Landau} L. D. Landau, J. Phys. (U.S.S.R), {\bf  5}, 71, (1941).


\bibitem{Feyn54}R. P. Feynman, Phys. Rev. {\bf 94}, 262 (1954).


\bibitem{Eric1}E. Akkermans and R. Narevich, Phil Mag. B, {\bf 77}, 1097
(1998).

\bibitem{ericlh}E. Akkermans and K. Mallick, in {\it Topological 
aspects of low dimensional systems}, 
Les Houches Summer School, Session 
{\bf LXIX} (Springer, Berlin 1999); cond-mat/9907441.  \S3.5
is particularly relevent for the present discussion. 

\bibitem{Tricomi}F. G. Tricomi, Fonctions Hyperg$\acute{e}$ometriques
Confluentes, M$\acute{e}$morial 
des Sciences Math$\acute{e}$matiques, CXL, (1960).

\bibitem{AS}M. Abramowitz and  I. A. Stegun,
{\it Handbook of Mathematical Functions}, 
(Dover Publications, New York, 1968).   

\bibitem{Eric11}E. Akkermans, J. E. Avron, R. Narevich and R. Seiler,   
The Eur. Phys. J. B {\bf 1}, 117-121, (1998).

\bibitem{Eric2} E.Akkermans and R. Narevich ({\it unpublished}).

\bibitem{Narevich} R. Narevich, {\it Ph.D.thesis}, 
Technion I.I.T,Israel
({\it unpublished})

\bibitem{Halperin} B. I. Halperin, Phys. Rev. B. {\bf 25}, 2185, (1982.)

\bibitem{APS} M. Atiyah, V. Patodi, I Singer, 
{\it Math. Proc. Camb. Phys. Soc.}, {\bf 77}, 43, (1975).


\bibitem{Peeters}A. K. Geim {\it et. al.}, Nature, {\bf 390}, 
259 (1997), P. Singh. Deo, V. A. Schweigert, F. M. Peeters, 
and A. K. Geim, Phys. Rev. Lett., {\bf 79}, 4653 (1997).

\bibitem{Eric3} E. Akkermans and K. Mallick, J. Phys. A {\bf 32}, 7133
(1999) and E. Akkermans, D. M. Gangardt and K. Mallik, Phys. Rev. B {\bf 62},
12427, (2000).

\bibitem{Lopez} C. Bolech, G. C. Buscaglia and A. L$\acute{o}$pez, 
Phys. Rev. B {\bf 52}, R15179 (1995).  

\bibitem{LSC04}C. Lobo, A. Sinatra and Y. Castin, Phys. Rev. Lett, 
{\bf 92}, 020403 (2004).

\bibitem{Math} S. Wolfram, 
{\it The MATHEMATICA BOOK}, (Cambridge University Press, 1996), 
Third Edition. 
\end{thebibliography}
\end{document}